\documentclass[sigconf,anonymous=false,review=false]{acmart}

\renewcommand\footnotetextcopyrightpermission[1]{} 

\AtBeginDocument{%
  \providecommand\BibTeX{{%
    \normalfont B\kern-0.5em{\scshape i\kern-0.25em b}\kern-0.8em\TeX}}}

\setcopyright{acmcopyright}
\copyrightyear{2023}
\acmYear{2023}
\acmDOI{nnnnn}

\acmConference[Under review]{ }{2023}{Conference}
\acmBooktitle{ Under review}
\acmPrice{15.00}
\acmISBN{nnn}

\acmSubmissionID{4527}

\usepackage{booktabs} 
\usepackage{diagbox} 
\usepackage{graphicx}
\usepackage[caption=false]{subfig}
\usepackage{color}
\usepackage{microtype}
\usepackage{amsmath,amsfonts}
\usepackage{mathtools}  
\usepackage{mathrsfs} 
\usepackage{bm}
\usepackage{booktabs, multirow}  
\usepackage{makecell}
\usepackage{stfloats}
\usepackage[thicklines]{cancel}
\usepackage[ruled,linesnumbered]{algorithm2e}

\begin{document}

\title{Gradient Coordination for Quantifying and Maximizing Knowledge Transference in Multi-Task Learning}

\author{Xuanhua Yang, Jianxin Zhao, Shaoguo Liu$^{\scriptscriptstyle *}$, Liang Wang and Bo Zheng}
\thanks{${\scriptscriptstyle *}$ Correspondence to: Shaoguo Liu.}
\affiliation{
  \institution{Alibaba Group, Beijing, China} 
  \country{}
}
\email{{xuanhua.yxh,zhaojianxin.zjx,shaoguo.lsg,	liangbo.wl,bozheng}@alibaba-inc.com}

\begin{abstract}
Multi-task learning (MTL) has been widely applied in online advertising and recommender systems. To address the \textit{negative transfer} issue, recent studies have proposed optimization methods that thoroughly focus on the gradient alignment of directions or magnitudes. However, since prior study has proven that both general and specific knowledge exist in the limited shared capacity, overemphasizing on gradient alignment may crowd out task-specific knowledge, and vice versa. In this paper, we propose a transference-driven approach \textbf{CoGrad} that adaptively maximizes knowledge transference via \textbf{Co}ordinated \textbf{Grad}ient modification. We explicitly quantify the transference as loss reduction from one task to another, and then derive an auxiliary gradient from optimizing it. We perform the optimization by incorporating this gradient into original task gradients, making the model automatically maximize inter-task transfer and minimize individual losses. Thus, \textsf{CoGrad} can harmonize between general and specific knowledge to boost overall performance. Besides, we introduce an efficient approximation of the Hessian matrix, making \textsf{CoGrad} computationally efficient and simple to implement. Both offline and online experiments verify that \textsf{CoGrad} significantly outperforms previous methods.
\end{abstract}

\maketitle
\fancyhead{}

\section{Introduction}\label{intro}

\begin{figure}[t]
\centering
\centerline{\includegraphics[width=1.0\columnwidth]{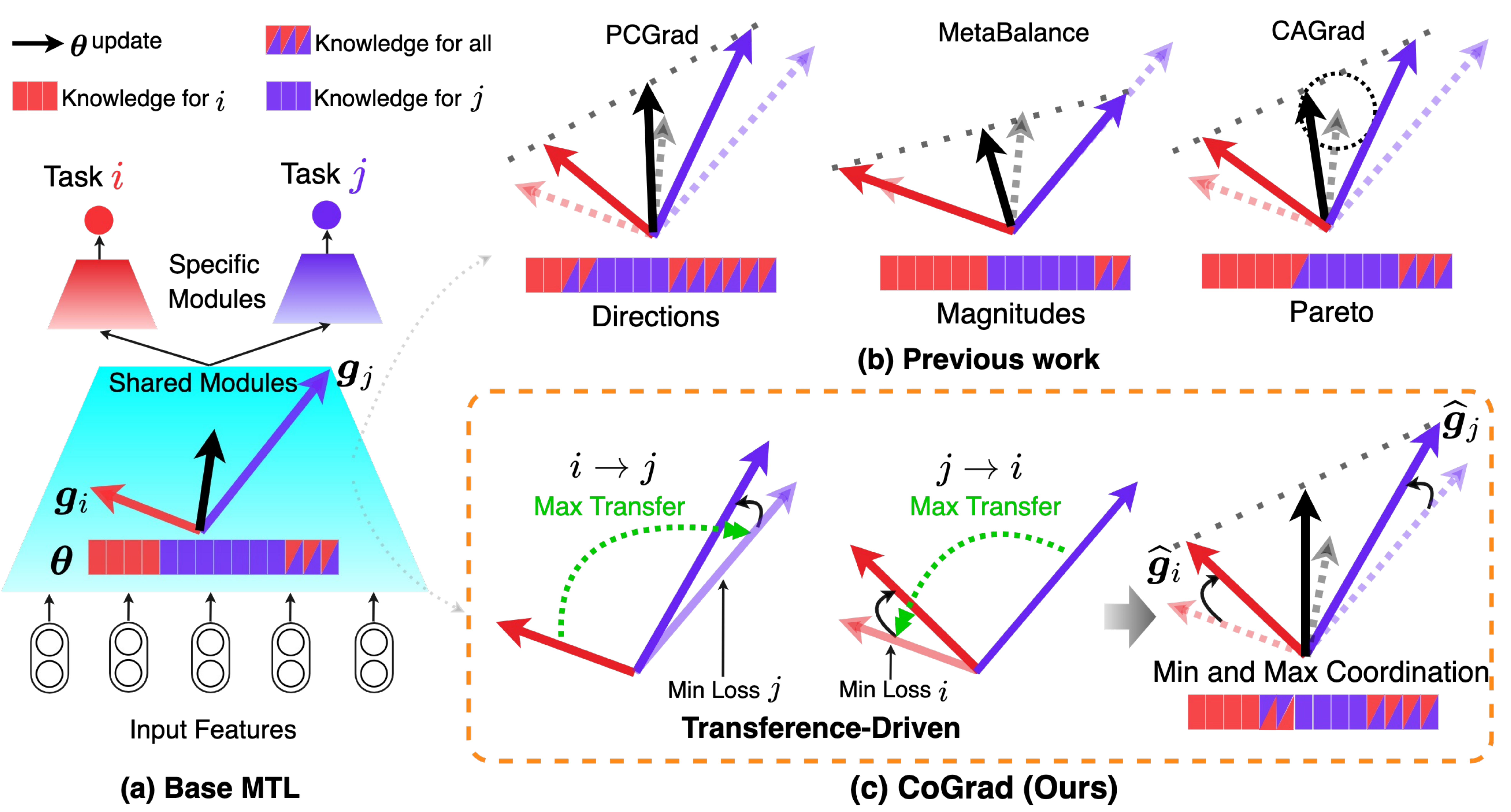}}
\vspace{-1em}
\caption{(a) shows MTL paradigm and the base gradient decent. Compared to previous work (b) that homogenize gradient directions or magnitudes, \textsf{CoGrad} (c) modifies gradients using a transference-driven way that maximizes inter-task transfer while simultaneously minimizes individual losses.}
\label{fig:overall}
\vspace{-1em}
\end{figure}

In online advertising and recommender systems, multi-task learning (MTL) has been proven to be extremely effective to simultaneously predict multiple user behaviours (e.g., clicking, viewing and buying)~\cite{wen2021hierarchically,wei2021autoheri}. As Fig.~\ref{fig:overall} (a) shown, popular backbones of MTL involve shared modules for encoding multi-behaviour representations and several specialized heads to output task-specific predictions. In contrast to single-task learning (STL), shared modules are regarded as the key components for reciprocal benefits across tasks.

\begin{figure}[t]
\centering
\centerline{\includegraphics[width=1.0\columnwidth]{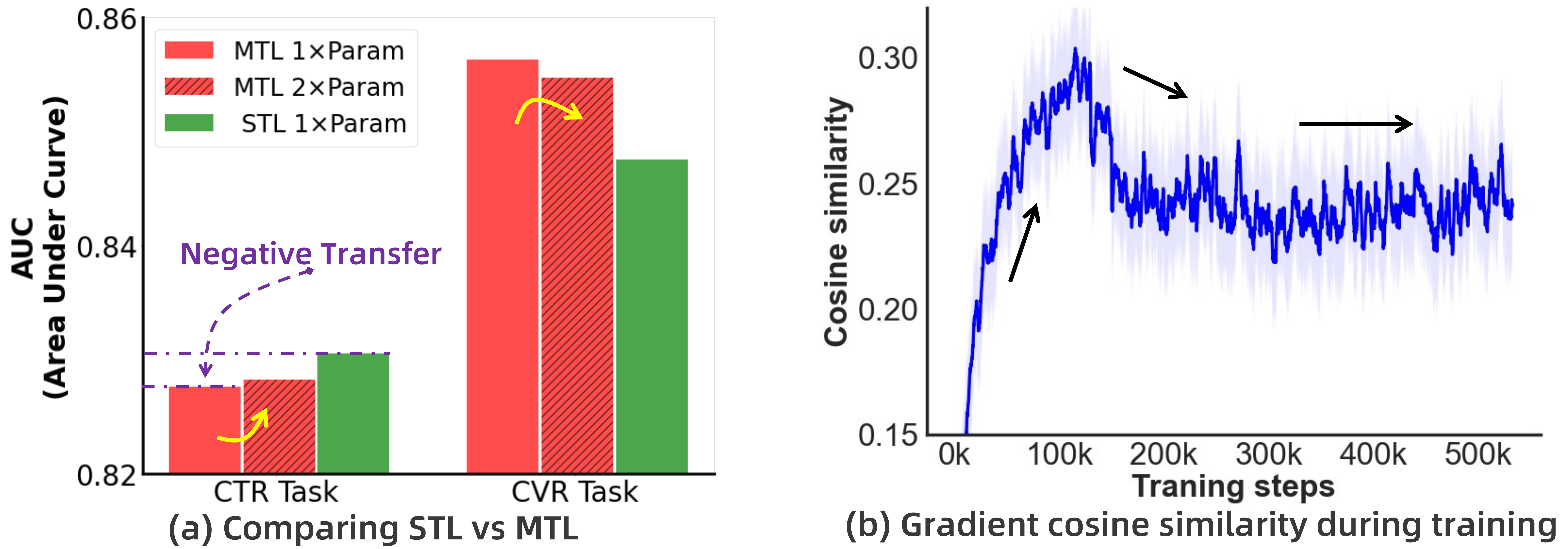}}
\vspace{-1.0em}
\caption{(a) Negative transfer problem and effects of expanding model capacity. (b) Trend of gradient cosine similarity.}
\label{fig:introduction}
\vspace{-1.5em}
\end{figure}

The most crucial issue induced by encoding general knowledge is \textit{negative transfer}~\cite{ruder2017overview}. 
As seen in Fig.~\ref{fig:introduction}(a), compared to STL, the CTR\footnote{CTR: click-through rate, CVR: post-click conversion rate} task performance  decreases (i.e.,\textit{negative transfer}) while the CVR task achieves improvement. The similar phenomena are also observed in~\cite{lin2022personalized,wang2020negative}. Recently, several optimization methods that modify task gradients based on gradient alignment strategies have been proposed to address this issue. \textsf{PCGrad}~\cite{2020pcgradient} and \textsf{GradVac}~\cite{2020gradvac} modify the gradient directions to maintain the directional consistency to enhance transference.
\textsf{GradNorm}~\cite{2018gradnorm} and \textsf{MetaBalance}~\cite{he2022metabalance} homogenize gradient magnitudes to prevent shared modules from being dominated by certain tasks of larger gradient magnitudes. 
\textsf{MGDA}~\cite{MGDA} and \textsf{CAGrad}~\cite{2021cagrad} utilize \textit{Pareto} solution to manipulate both the directions and magnitudes to mitigate gradient conflict. 

However, since prior study has demonstrated that both general and specific knowledge exists in a competitive manner in shared modules~\cite{wang2020negative}, overemphasizing on the gradient alignment may crowd out the specific knowledge that is also useful for individual task. For example, direction-based methods usually align gradients by raising the cosine similarity, as it is regarded to be highly related with transference~\cite{2020pcgradient}. Yet, this could even degrade overall performance when dealing with two weakly correlated tasks. On the other hand, in our settings, we observed that the cosine similarity rises in early training but then diminishes (shown in Fig.~\ref{fig:introduction}(b)). This implies that MTL models are prone to encoding specific knowledge, which may in turn consume capacity budgets of general information.

Consequently, over-encoding either general or specific knowledge could decrease overall performance. To address this, a straightforward way is to expand the shared parameters. However, due to the model's tendency for specificity, this may make it more likely to learn specific rather than general knowledge. This is supported by the observation that expanding parameters increases CTR performance but decreases CVR in Fig.~\ref{fig:introduction}(a). Therefore, it is challenging to coordinate the encoding of general and specific knowledge.

In this paper, we propose a transference-driven approach \textbf{CoGrad} that adaptively maximizes knowledge transference via \textbf{Co}ordinated \textbf{Grad}ient modification. Specifically, we theoretically quantify the inter-task transfer as the loss reduction of one task induced by the update from another task gradient. Then, we optimize this quantification to derive an auxiliary gradient, and incorporate it into original task gradients. In this way, \textsf{CoGrad} can maximize inter-task transfer while simultaneously minimize individual losses. Thus, \textsf{CoGrad} achieves the harmonization between general and specific knowledge, improving overall performance.
Besides, \textsf{CoGrad} contains a Hessian matrix, resulting in expensive computations. We additionally introduce an efficient Hessian matrix approximation to make \textsf{CoGrad} computationally efficient and simple to implement in industrial applications.
Our contributions are:
\begin{itemize}
    \item To our knowledge, we are the first to explicitly quantify inter-task transfer and utilize it for gradient modulation, which has promising applications for MTL.
    \item We propose a transference-driven gradient modulation method (\textsf{CoGrad}) that can adaptively maximize inter-task transfer, which is also computationally efficient. 
    \item Experiments show that \textsf{CoGrad} outperforms prior baselines. Our empirical analysis verifies that \textsf{CoGrad} can effectively harmonize general and specific knowledge.
\end{itemize}

\section{Problem Formulation}
In multi-task recommendation, considering a set of $T$ tasks $\mathcal{T} = \{t\}_{t=1}^T$, and the training dataset $\mathcal{D} = \{(x_n,\{y_n^t\}_{t \in \mathcal{T}})\}^{|\mathcal D|}_{n=1}$, where $x_n$ and $y_n^t$ represent the feature vectors (including user features, item features and context features) and binary user feedback label (whether or not a user has clicked, viewed, bought, etc.) of $n_{th}$ instance, respectively. We denote the $t_{th}$ task dataset as $\mathcal{D}_t=\{x_n,y_n^t\}^{|\mathcal D|}_{n=1}$.

Let $\bm L_t(\mathcal{D}_t;{\theta},{\phi}_t)$ denote the loss on $\mathcal{D}_t$ for task $t$, where $\theta$ and $\phi_t$ represent shared and specific parameters, respectively. The standard MTL loss is aggregated as a weighted sum formulation:
\begin{equation}
    \mathcal{L}(\mathcal{D};\theta,\{\phi_t\}_{t \in \mathcal{T}}) = \begin{matrix} \sum_{t \in \mathcal{T}} \end{matrix} w_t \bm L_t(\mathcal{D}_t;{\theta},{\phi}_t)
\end{equation}
Here, we omitted the regularization term and $\{w_t\}_{t=1}^T$ represent weights for scaling task losses. All parameters are updated as:
\begin{equation}\label{eq:task-sharing}
    \theta^{k+1}=\theta^k-\eta \begin{matrix} \sum_{t \in \mathcal{T}} \end{matrix}w_t\nabla_{\theta} \bm L_t(\mathcal{D}_t;{\theta}^k,{\phi}_t^k)
\end{equation}
\begin{equation}\label{eq:task-specific}
    \phi_t^{k+1}=\phi_t^k-\eta w_t \nabla_{\phi_t}\bm L_t(\mathcal{D}_t;{\theta}^k,{\phi}_t^k), \forall t \in \mathcal{T}
\end{equation}

MTL achieves knowledge transference by iterating Eq.~\ref{eq:task-sharing}. Instead of modifying the gradient magnitudes or directions as most prior methods, we attempt to explicitly quantify the inter-task transfer to guide the optimization in Eq.~\ref{eq:task-sharing}.

\section{Proposed Approach}
During iterating $\theta$ , we expect that the gradient update from one task will minimize its own loss as well as help reduce the loss of another task as much as possible. We achieve this challenging goal via quantifying and maximizing knowledge transference to perform our optimization. We will elaborate our approach in this section. For simplicity, we use $\bm{L}_t(\theta)$ and $\bm{g}_t(\theta)$ to denote $\bm{L}_t(\mathcal{D}_t;{\theta},{\phi}_t)$ and $\nabla_{\theta} \bm{L}_t(\mathcal{D}_t;{\theta},{\phi}_t)$ respectively.

\begin{algorithm}[t]
    \small
    \caption{Training Algorithm for \textsf{CoGrad}}
    \label{alg:training}
    \KwIn{Training dataset $\mathcal D$, initial parameters $\{\theta\} \cup \{\phi_i \}_{i \in \mathcal{T}}$, learning rate $\eta$, $\{w_i\}_{ i \in \mathcal{T} }$ for scaling task losses, $\{\gamma_i\}_{i \in \mathcal{T}}$ for controlling the strength of maximizing transference.
    } 
    \While{Not converged}
    {
    Sample a batch of samples $\mathcal B$ from $\mathcal D$\;
    Update task-specific parameters:\\
    $\phi^{k+1}_i = \phi^k_i - \eta \nabla_{\phi_i} \bm L_i(\mathcal{B}; \theta^k, \phi_i^k), \forall i \in \mathcal{T}$\;
    \textbf{CoGrad optimization for shared parameters}:\\
      Compute each task gradient on $\theta$:\\
      $\bm g_i( \theta^{k} ) = \nabla_{\theta} \bm L_i(\mathcal{B}; \theta^k, \phi_i^k), \forall i \in \mathcal{T}$\;
      Compute \textbf{CoGrad}, $\forall i \in \mathcal{T}$:\\
      $\widehat{\bm g}_i(\theta^k) = \bm g_i(\theta^k) - \begin{matrix} \sum_{ j \neq i, j \in \mathcal{T}} \end{matrix}\gamma_j \bm g_i(\theta^k) \odot \bm g_i(\theta^k)\odot \bm g_j(\theta^k)$\;
      Update shared parameters $\theta$:\\
      $\theta^{k+1} = \theta^k - \eta \begin{matrix} \sum_{ i \in \mathcal{T}} \end{matrix} w_i \widehat{\bm g}_i(\theta^k) $\;
    }
\end{algorithm}

\subsection{Quantifying Knowledge Transference}\label{method:inter-task-impacts}

Quantifying knowledge transference in MTL has huge potential to enhance generalization. Prior work
~\cite{Fifty2021EfficientlyIT} has measured the inter-task affinity for grouping tasks. Inspired by this, we define the quantification of transfer from task $i$ to $j$ ($i,j \in \mathcal{T}$) as the loss reduction of task $j$ induced by the update from task $i$ gradient. Specifically, assuming that the shared parameters $\theta$ is updated by task $i$ at time-step $k$ with learning rate $\gamma_i > 0$, we have:
\begin{equation}\label{eq:virtual-theta}
    \theta^{k+\tau_i}=\theta^k-\gamma_i \bm{g}_i(\theta^k)
\end{equation}
Then, we use $\theta^{k+\tau_i}$ to examine the impacts on task $j$ loss by comparing the loss changes of task $j$ before and after, formulated as:
\begin{equation}\label{eq:inter-task impact}
\begin{aligned}
    \Delta^k \bm{L}_{i \to j}
    & = \bm{L}_j(\theta^k) - \bm{L}_j(\theta^{k+\tau_i})
\end{aligned}
\end{equation}
Plugging Eq.\eqref{eq:virtual-theta} into Eq.\eqref{eq:inter-task impact} and making a first-order Taylor series expansion of $\bm L_j(\theta^{k+\tau_i})$ yields:
\begin{equation}\label{eq:inter-task inner product}
\begin{aligned}
    \Delta^k \bm{L}_{i \to j}
    & = \bm{L}_j(\theta^k) - \bm{L}_j(\theta^k-\gamma_i \bm{g}_i(\theta^k))  \approx \gamma_i \bm{g}_i^T( \theta^k ) \bm{g}_j(\theta^k)
\end{aligned}
\end{equation}

Notice that, $\bm{g}_i^T( \theta^k ) \bm{g}_j(\theta^k)$ shows that larger inner product value means better transference, which may explain the observation
in Fig.~\ref{fig:introduction}(b) that cosine similarity has a tendency to increase in the early training. Eq.~\ref{eq:inter-task inner product} also implies that improving gradient inter product can enhance transference. This is relatively consistent with previous work~\cite{2020pcgradient,2020gradvac,ruder2017overview}, justifying our quantification. Since $\Delta^k \bm L_{i \to j}$ reflects the impacts from task $i$ to $j$, it can be regarded as the quantification of inter-task ($i \to j$) transfer. Next, we introduce how to use this quantification to perform a coordinated optimization that can harmonize general and specific knowledge.

\subsection{Coordinated Gradient Modulation}\label{method:max inter-task}
\subsubsection{\textbf{Maximizing Inter-Task Transfer}}
We maximize the inter-task transfer $\Delta^k \bm L_{i \to j}$ (Eq.~\ref{eq:inter-task impact}) to derive the optimization gradient. Here, we fix task $i$ since we merely consider the impacts from task $i$ to $j$ (i.e., approximating $\nabla_\theta \theta^{k+\tau_i}$ as $1.0$) and yield:
\begin{equation}\label{eq:origin inter gradient}
    \nabla_{\theta} \Delta^k \bm L_{i \to j} = \bm g_j(\theta^k) - \bm g_j(\theta^{k+\tau_i})
\end{equation}
Making a first-order Taylor series expansion of $\bm g_j(\theta^{k+\tau_i})$ yields:
\begin{equation}\label{eq:final inter gradient}
\begin{aligned}
    \nabla_{\theta} \Delta^k \bm L_{i \to j} 
    & = \bm g_j(\theta^k) - \left( \bm{g}_j(\theta^k) - \gamma_i \bm H_j(\theta^k) \bm{g}_i(\theta^k) \right ) \\
    & = \gamma_i \bm H_j(\theta^k) \bm g_i(\theta^k) \\
\end{aligned}
\end{equation}
where $\bm H_j(\theta^k)$ is the Hessian matrix of $\bm L_j(\theta^k)$.
\subsubsection{\textbf{General and Specific Knowledge Harmonization}}

Here, $-\nabla_{\theta} \Delta \bm{L}^k_{i \to j}$ can represent the gradient for maximizing transference from task $i$ to $j$. Then, we incorporate it into the original gradient $\bm g_j(\theta^k)$ (See Eq.~\ref{eq:gradient balance j}), leading to both maximizing transference from task $i$ to $j$ (second term in Eq.~\ref{eq:gradient balance j}) and minimizing the individual loss of task $j$ (first term in Eq.~\ref{eq:gradient balance j}), as illustrated in Fig.~\ref{fig:overall}(c):
\begin{equation}\label{eq:gradient balance j}
\begin{aligned}
    \widehat{\bm g}_j(\theta^k) 
    & = \bm g_j(\theta^k) - \gamma_i \bm H_j(\theta^k) \bm g_i(\theta^k) \\
\end{aligned}
\end{equation}
where the hyper-parameter $\gamma_i$ can be regarded as the balance degree between maximizing transference and minimizing specific losses. In fact, \textsf{CoGrad} is insensitive to $\gamma_i$ as discussed in section~\ref{method:hyper}. We can obtain the counterpart for task $i$ in the same way. Based on this, we introduce a more general formulation as:
\begin{equation}\label{eq:gradient balance i}
\begin{aligned}
    \widehat{\bm g}_i(\theta^k) 
    & = \bm g_i(\theta^k) - \begin{matrix} \sum_{ j \neq i, j \in \mathcal{T}} \end{matrix}\gamma_j \bm H_i(\theta^k) \bm g_j(\theta^k), \forall i \in \mathcal{T}\\
\end{aligned}
\end{equation}

\subsection{Maximizing Transference Approximation}\label{method:approximation}
Due to the Hessian matrix $\bm H_j(\theta^k)$ in Eq.~\ref{eq:final inter gradient}, calculating the gradient 
$\nabla_\theta \Delta^k \bm L_{i \to j}$ for maximizing transference is too expensive both in storages and computations. Alternatively, we introduce an applicable approximator in ~\cite{zheng2017approximation} (see Eq.~\ref{eq:approximation}), where the approximation accuracy has been theoretically and empirically guaranteed.
\begin{equation}\label{eq:approximation}
    \bm H_j(\theta^k) \bm g_i(\theta^k) = \lambda^k \bm g_j(\theta^k) \odot \bm g_j(\theta^k)\odot \bm g_i(\theta^k)
\end{equation}
where $\odot$ is \textit{Hadamard product} (i.e., element-wise product) and $\lambda^k$ is a hyper-parameter.  According to ~\cite{zheng2017approximation}, we fix $\lambda^k$ to 1.0 in our method. We plug Eq.~\ref{eq:approximation} into Eq.~\ref{eq:gradient balance i}, leading to a simple and efficient training process of \textsf{CoGrad} as elaborated in Alg.~\ref{alg:training}.

\subsection{Discussion}\label{method:discussion}
\subsubsection{\textbf{Connections with Meta-Learning}}\label{method:connections-metalearning}\textsf{Sequential Reptile} (\textsf{Seq.Rept})~\cite{lee2021seqReptile} resembles our method in terms of a balance in general and specific information. \textsf{Seq.Rept} aligns task gradients using meta-learning based on inner-loops trajectory with all tasks sequentially. Informally, \textsf{Seq.Rept} is very similar to  \textsf{CoGrad} under conditions with two tasks and two inter-steps. These connections also help the comprehension of our approach. However, when facing with more tasks, \textsf{Seq.Rept} becomes computationally expensive due to the inter and outer loops, while \textsf{CoGrad} remains efficient with negligible computation increase. Moreover, it is also business-sensible to explicitly quantify and  maximize inter-task transfer. Thereby, \textsf{CoGrad} is more practical in industrial applications.

\subsubsection{\textbf{Hyper-Parameters Selection}}\label{method:hyper}
Our method introduces new hyper-parameters $\{ \gamma_j\}_{j \in \mathcal{T}}$ in Eq.~\ref{eq:gradient balance i}, which control the strength of maximizing inter-task transfer during optimization iterations. From our theoretical derivation, we can regard $\gamma_j$ as the virtual learning rate for updating parameters of task $j$. Thus, selecting a reasonably tiny value suffices. Moreover, We also empirically find that the performance is insensitive to the hyper-parameters, and they only have a small impact on convergence speed.

\section{Experiments}

\subsection{Experimental Setup}
\subsubsection{\textbf{Datasets}} We conduct the offline experiments on two industrial datasets. The first is a public dataset named \textbf{Ali-CCP}~\cite{ma2018entire}, which contains behaviors of clicking and buying. The second dataset including other another conversion behavior (i.e., viewing commodity details page) is collected from our large-scale E-commerce adverting platform, named \textbf{Ecomm}. We split each dataset into training/validation/test sets by timestamp with 4:1:1 proportion. Table~\ref{statistics} lists the statistics.

\begin{table}[tb]
\small
\caption{Statistics of two datasets.}
\vspace{-1.3em}
\centering
\begin{tabular}{ccccc}
\toprule
\textbf{Dataset} & \# impression &  \# click & \# view & \# buy  \\
\midrule
Ali-CCP & 84 mil. &  3.4 mil. & -- & 18 k \\
Ecomm & 1.6 bil. &  25mil.  & 10 mil. & --\\
\bottomrule
\end{tabular}
\label{statistics}
\vspace{-1.0em}
\end{table}

\subsubsection{\textbf{Competitors}} 
We compare \textsf{CoGrad} with the previous state-of-the-art gradient modulation techniques by using \textsf{Shared Bottom} ~\cite{caruana1997multitask} and \textsf{MMOE}~\cite{Mmoe} as two fundamental MTL architectures: 
(1) \textsf{PCGrad}~\cite{2020pcgradient} aggressively projects task gradients based on  directional consistency. 
(2) \textsf{MetaBalance}~\cite{he2022metabalance} (\textsf{MBalance} for short) adaptively homogenizes gradient magnitudes to prevent model from being dominated by certain tasks.
(3) \textsf{CAGrad}~\cite{2021cagrad} searches around the average gradients to maximize the worst task performance.
(4) \textsf{Seq.Rept}~\cite{lee2021seqReptile} uses inner-loops trajectory with sequential tasks to maximize gradient alignment. 

We use AUC (Ali-CCP) and Group AUC of users (an industrial metric in ~\cite{he2016ups} for Ecomm) as the evaluation metrics. To ensure fair comparison, the feature embedding size is fixed to 8 for all methods (including STL) and each network contains three hidden layers with \{512,256,128\} for Ecomm and \{128,64,32\} for Ali-CCP. The first two layers are shared and the last is specific in MTL models. And for MMOE, we set three experts. We use Adam~\cite{kingma2014adam} optimizer with 1024 (Ecomm) and 256 (Ali-CCP) batch size and 0.001 (Ecomm) and 0.005 (Ali-CCP) learning rate. The loss scaling weights are set by a heuristic way based on priori statistical cross-entropy (excluding \textsf{MBalance}). Other hyper-parameters in baselines are carefully tuned based on original researches. Our hyper-parameters $\{\gamma_{ctr}, \gamma_{cvr}\}$ are set to $\{0.003,0.001\}$ (Ecomm) and $\{0.01,0.005\}$ (Ali-CCP).

\subsection{Results and Discussion}
\subsubsection{\textbf{Main Results}} Table~\ref{results:main} shows the results of all methods on production and public datasets. For both datasets, all MTL methods outperform \textsf{Single DNN} on the CVR task while perform slightly worse on the CTR task. This is because that the \textit{data sparsity} problem~\cite{ma2018entire} makes it difficult to fit a single CVR model, but the CTR can learn effectively due to the adequate data. Compared to \textsf{Shared Bottom}-based methods, \textsf{MMOE}-based ones perform marginally better on the CTR but worse on the CVR. This is consistent with our perspective that the explicit CTR-specific experts help encode more CTR-specific knowledge, but take over the capacity budgets for encoding CVR-specific or general knowledge.

\textsf{CAGrad}, which focuses on the worst task, and \textsf{Seq.Rept}, which considers individual tasks, outperform previous methods towards preventing negative transfer. This implies that general and specific knowledge are equally important in shared modules. Benefiting from the transference-driven technique, which harmonizes both general and specific knowledge, \textsf{CoGrad} significantly surpasses all baselines on the CVR, and meanwhile, achieves comparable CTR performance compared to STL.

Notice that, compared to Seq.Rept, the improvement of \textsf{CoGrad} seems minor on Ecomm dataset. Firstly, achieving industrial 0.1 AUC gain is remarkable~\cite{ma2018entire}. Secondly, as discussed in section~\ref{method:connections-metalearning}, \textsf{CoGrad} resembles \textsf{Seq.Reqt} in two-task settings. However, \textsf{CoGrad} is more efficient in computation and simple to implement, both of which are crucial in large-scale applications.

\begin{table}[t]
\footnotesize
\vspace{-1.0em}
\caption{Average results (five runs) on two datasets where \textbf{bold} and \underline{underline} represent the best and runner-up respectively. ``*'' denotes the improvement significance at the level of $p<0.05$. $\Delta$ denotes performance gain w.r.t. \textsf{Single DNN}}
\vspace{-1.3em}
\centering
\setlength{\tabcolsep}{1.2mm}{
\begin{tabular}{lcccc}
\toprule
\multirow{2}*{\textbf{Approach}} & \multicolumn{2}{c}{\textbf{Ecomm} (production)}&
\multicolumn{2}{c}{\textbf{Ali-CCP} (public)} \\
\cmidrule(lr){2-3}\cmidrule(lr){4-5}
 & GAUC$_\mathrm{ctr}$($\Delta$) & GAUC$_{\mathrm{cvr}}$($\Delta$) & AUC$_\mathrm{ctr}$($\Delta$) &  AUC$_\mathrm{cvr}$($\Delta$)  \\   
\midrule
\textsf{Single DNN} & \textbf{76.24} & 78.11 & \underline{63.97}  & 65.85 \\
\textsf{Shared Bottom} & 76.01 ($\downarrow$0.23) & 78.97 ($\uparrow$0.86) & 63.81 ($\downarrow$0.16) & 66.71 ($\uparrow$0.86) \\
\textsf{\qquad+PCGrad}& 76.07 ($\downarrow$0.17) & 79.21 ($\uparrow$1.10) & 63.84 ($\downarrow$0.13) & 67.25 ($\uparrow$1.40)  \\
\textsf{\qquad+MBalance}& 76.04 ($\downarrow$0.20) & 78.92 ($\uparrow$0.81) & 63.78 ($\downarrow$0.19) & 66.99 ($\uparrow$1.14)  \\
\textsf{\qquad+CAGrad}& 76.08 ($\downarrow$0.16) & 79.34 ($\uparrow$1.13) & 63.80 ($\downarrow$0.17)& 67.38 ($\uparrow$1.53) \\
\textsf{\qquad+Seq.Rept} & 76.10 ($\downarrow$0.14) & \underline{79.50 ($\uparrow$1.39)} & \textbf{63.98 ($\uparrow$0.01)} & \underline{67.56 ($\uparrow$1.71)}  \\

\cmidrule(lr){1-1}
\textsf{\qquad+CoGrad}& \underline{76.17 ($\downarrow$0.07)} & \textbf{79.61 ($\uparrow$1.50) }$^*$ & \underline{63.97 ($\downarrow$0.00) } & \textbf{67.78 ($\uparrow$1.93)} $^*$ \\

\cmidrule(lr){1-5}
\textsf{MMOE}& 76.05 ($\downarrow$0.19) & 78.93 ($\uparrow$0.82) & 63.84 ($\downarrow$0.13) & 66.49 ($\uparrow$0.64)  \\
\textsf{\qquad+PCGrad}& 76.07 ($\downarrow$0.17) & 79.13 ($\uparrow$1.02) & 63.89 ($\downarrow$0.08) & 66.61 ($\uparrow$0.76)  \\
\textsf{\qquad+MBalance}& 76.03 ($\downarrow$0.21) & 78.86 ($\uparrow$0.75) & 63.85 ($\downarrow$0.12) & 66.39 ($\uparrow$0.54)  \\
\textsf{\qquad+CAGrad}& 76.10 ($\downarrow$0.14) & 79.33 ($\uparrow$1.22) & 63.91 ($\downarrow$0.06) & 66.85 ($\uparrow$1.00)  \\
\textsf{\qquad+Seq.Rept}& 76.11 ($\downarrow$0.13) & \textbf{79.45 ($\uparrow$1.34) } & \underline{63.93 ($\downarrow$0.04)} & \underline{67.15 ($\uparrow$1.30)} \\

\cmidrule(lr){1-1}
\textsf{\qquad+CoGrad}& \underline{76.21 ($\downarrow$0.03)} & \underline{79.42 ($\uparrow$1.31)} & \textbf{63.96 ($\downarrow$0.01)} & \textbf{67.32 ($\uparrow$1.47) }$^*$ \\
\bottomrule
\end{tabular}}
\label{results:main}
\vspace{-1.0em}
\end{table}

\subsection{Further Analysis}

\subsubsection{\textbf{Robustness on capacity size}}
We double the size of first hidden layer to examine the robustness w.r.t. the capacity size with \textsf{Shared Bottom}. In base model, expanding capacity tends to mitigate negative transfer on the CTR at the cost of hurting the CVR (shown in Table~\ref{table:robustness}). \textsf{CoGrad}, however, improves CTR performance while maintaining CVR unaffected. This verifies that \textsf{CoGrad} is strongly robust that can adaptively maximize knowledge transference.

\subsubsection{\textbf{Visualization of general and specific knowledge harmonization}}
\begin{figure}[t]
\vspace{-1.0em}
\centering
\centerline{\includegraphics[width=1.0\columnwidth]{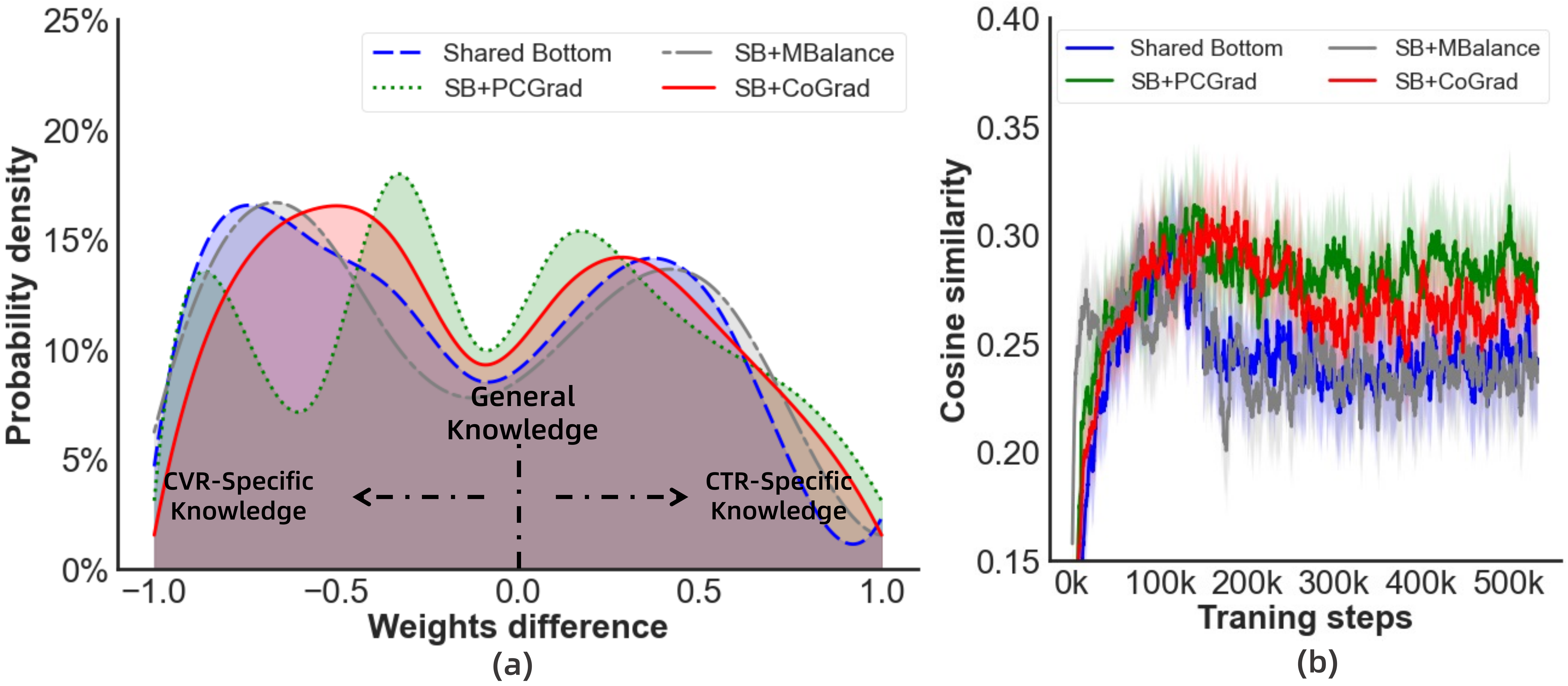}}
\vspace{-1.0em}
\caption{(a) Distribution of the importance weights difference w.r.t hidden units in the last shared layer. (b) Comparison between \textsf{CoGrad} and baselines on gradient similarity.}
\label{fig:further_study}
\vspace{-1em}
\end{figure}

We investigate the ability of general and specific knowledge harmonization with the help of the \textit{pretrain-finetune} paradigm. We freeze the total shared parameters of the trained model and connect the last shared layer to a trainable linear layer. We then fine-tune this linear layer for each task to learn the hidden units weights of the last shared layer (128 dimensions). The intuition behind this is that if both tasks consider one unit to be important, their normalized weights should be close, and vice versa. Ultimately, We compute the weights (normalized) difference (i.e., CTR weights minus CVR weights), leading to a distribution (smoothed) as illustrated in Fig.~\ref{fig:further_study}(a). The knowledge is considered to be more general when the difference is closer to zero, and more specific when the difference is further away from zero.

First, this visualization demonstrates that the shared modules indeed contain general and specific knowledge, which is consistent with~\cite{wang2020negative}. Second, the area on the right side of zero (i.e., important knowledge for CTR task) is slightly smaller than that on the left side. This provides an explanation for the observation that CVR gained significantly from MTL while CTR was negatively affected. Third, the direction-based method (\textsf{PCGrad}) indeed increases the general knowledge by enforcing the gradient alignment, yet hinders the encoding of some specific knowledge. This supports our perspective described in the introduction that enforcing the gradient alignment may crowd out specific knowledge. Ultimately, compared to base MTL, \textsf{CoGrad} improves the general knowledge while it also maintains adequate specific knowledge. This verifies that \textsf{CoGrad} can effectively harmonize general and specific knowledge.

\begin{table}[t]
\footnotesize
\vspace{-1.5em}
\centering
\caption{Effects of varying hidden layer sizes on Ecomm. $\Delta$ denotes performance gain w.r.t. size $512 \times$.}
\vspace{-1.0em}
\centering
\setlength{\tabcolsep}{1.0mm}{
\begin{tabular}{lcccc}
\toprule
\multirow{2}*{\textbf{Methods}} & \multicolumn{2}{c}{GAUC$_\mathrm{ctr}$}&
\multicolumn{2}{c}{GAUC$_\mathrm{cvr}$} \\
\cmidrule(lr){2-3}\cmidrule(lr){4-5}
& Size $512 \times$ & Size $1024 \times$ ($\Delta$) & Size $512 \times$ & Size $1024 \times$ ($\Delta$) \\
\midrule
{\textsf{Shared Bottom}} & 76.01 & 76.05 ($\uparrow$0.04) & 78.97 & 78.81 ($\downarrow$0.16) \\
{\textsf{\quad+ \textsf{CoGrad}}} & 76.17 & \textbf{76.23 ($\uparrow$0.06)} & 79.61 & \textbf{79.63 ($\uparrow$0.02)} \\
\bottomrule
\end{tabular}}
\label{table:robustness}
\end{table}

\subsubsection{\textbf{Impacts on gradient similarity}} Fig.~\ref{fig:further_study}(b) shows the gradient cosine similarity of all methods during training. Compared to the base model, magnitude-based method (\textsf{Mbalance}) performs no improvement on the similarity, and direction-based method (\textsf{PCGrad}) achieves the most efficiency on gradient similarity. Considering the main results in Table~\ref{results:main}, we can validate that the cosine similarity is indeed related to the performance. However, their correlation is not exactly positive. \textsf{CoGrad} implicitly aligns gradients by the transference-driven technique, increasing cosine similarity to a certain level. Combining this with Fig.~\ref{fig:further_study}(a), we can confirm that \textsf{CoGrad} can automatically enhance the encoding of general knowledge to an appropriate degree, without over-encoding.

\subsection{Online A/B Test}
We conduct online experiments on our advertising system for 15 days compared to a well-trained MTL model.
We use metrics including $CTR=\frac{\text{\#click}}{\text{\#impression}}$, $CPC= \frac{\text{cost of advertisers}}{\text{\#click}}$, $CVR=\frac{\text{\#view}}{\text{\#click}}$,  and $CPA= \frac{\text{cost of advertisers}}{\text{\#view}}$. A higher CTR/CVR and a lower CPC/CPA indicate better performance. In large-scale industrial applications,  achieving 1\% gains is a significant improvement. \textsf{CoGrad} increases CTR and CVR by \textbf{2.03\%} and \textbf{4.75\%}, respectively, and reduces CPC and CPA by \textbf{1.64\%} and \textbf{5.23\%}. verifying its effectiveness in industry.

\section{Conclusion}
We propose \textsf{CoGrad}, a transference-driven approach that can automatically maximize inter-task transfer via coordinated gradient modification. It quantifies the transference and performs an optimization by maximizing this quantification and simultaneously minimizing task-specific losses, harmonizing both general and specific knowledge in shared modules to improve overall performance. Both offline and online experiments verify its effectiveness.

\bibliographystyle{ACM-Reference-Format}
\bibliography{research_src-InterGrad}

\end{document}